\documentclass[iop]{emulateapj}
\usepackage{graphics,graphicx,subfigure}
\usepackage{ textcomp }

\newcommand{\ee}[1]{\mbox{${} \times 10^{#1}$}}% scientific number format
\newcommand{\eten}[1]{\mbox{$10^{#1}$}}% power of ten
\newcommand{\msun}{\mbox{M$_\odot$}}% Msun
\newcommand{\degree}{\mbox{$^{\circ}$}}
\newcommand{\water}{H$_2$O}
\newcommand{\co}{$^{12}$CO}
\newcommand{\coo}{$^{13}$CO}
\newcommand{\cooo}{C$^{18}$O}
\newcommand{\hh}{\mbox{{\rm H}$_2$}}

\newcommand{\nnhp}{\mbox{\rm N}$_{2}$\rm{H}$^+$}

\begin{document}
\title {The Radial Distribution of \hh\ and CO in TW Hya as Revealed by Resolved ALMA Observations of CO Isotopologues}
\author{Kamber R. Schwarz\altaffilmark{1}, Edwin A. Bergin\altaffilmark{1}, L. Ilsedore Cleeves\altaffilmark{2}, Geoffrey A. Blake\altaffilmark{3}, Ke Zhang\altaffilmark{1}, Karin I. \"{O}berg\altaffilmark{2}, Ewine F. van Dishoeck\altaffilmark{4,5}, and Chunhua Qi\altaffilmark{2}}

\altaffiltext{1}{Department of Astronomy, University of Michigan, 1085 South University Ave., Ann Arbor, MI 48109, USA}
\altaffiltext{2}{Harvard-Smithsonian Center for Astrophysics, 60 Garden Street, Cambridge, MA 02138, SA}
\altaffiltext{3}{Division of Geological \& Planetary Sciences, MC 150-21, California Institute of Technology, 1200 E California Blvd, Pasadena, CA 91125}
\altaffiltext{4}{Leiden Observatory, Leiden University, P. O. Box 9513, 2300 RA Leiden, The Netherlands}
\altaffiltext{5}{Max-Planck-Institute f\"{u}r Extraterrestrische Physik, Giessenbachstrasse 1, Garching, 85748, Germany}

\begin{abstract}
CO is widely used as a tracer of molecular gas.
However, there is now mounting evidence that gas phase carbon is depleted in the disk around TW~Hya. 
Previous efforts to quantify this depletion have been hampered by uncertainties regarding the radial thermal structure in the disk. 
Here we present resolved ALMA observations of \coo~3-2, \cooo~3-2, \coo~6-5, and \cooo~6-5 emission in TW~Hya, which allow us to derive radial gas temperature and gas surface density profiles, as well as map the CO abundance as a function of radius. 
These observations provide a measurement of the surface CO snowline at $\sim$30~AU and show evidence for an outer ring of CO emission centered at 53~AU, a feature previously seen only in less abundant species. 
Further, the derived CO gas temperature profile constrains the freeze-out temperature of CO in the warm molecular layer to  $< 21$ K.
Combined with the previous detection of HD 1-0, these data constrain the surface density of the warm \hh\ gas in the inner $\sim30$~AU such that $\Sigma_{warm\,gas} = 4.7^{+3.0} _{-2.9} \mathrm{\,g\, cm^{-2}}\left(R/10\,\mathrm{AU}\right)^{-1/2}$. 
We find that CO is depleted by two orders of magnitude from $R=10-60$ AU,
with the small amount of CO returning to the gas phase inside the surface CO snowline insufficient to explain the overall depletion.
Finally, this new data is used in conjunction with previous modeling of the TW Hya disk to constrain the midplane CO snowline to 17-23~AU.
\end{abstract}


\section{Introduction}
It has long been thought that the primary carbon reservoir in protoplanetary disks is CO, as is the case for the ISM. While there is significant scatter from cloud to cloud, the CO abundance relative to \hh\ in warm molecular clouds is of order \eten{-4} \citep{Lacy94}. 
Lower CO abundances of order \eten{-6} have been inferred for the disks around several Herbig Ae and T Tauri stars, with the anomalously low abundance attributed to either photodissociation of CO, CO freeze-out onto grain mantles, grain growth, or a low total gas mass \citep{vanZadelhoff01,Dutrey03,Chapillon08}. %Chapillon: photodissociation, Dutrey: CO depletion, low gas-to-dust ratio

In order to determine the fractional abundance of CO, one must first determine the total disk mass, the majority of which resides in \hh, which does not readily emit at the relevant temperatures. CO, particularly the less abundance isotopologues \coo\ and \cooo, is often used as a tracer of the total gas mass. However, if the goal is to measure the fractional CO abundance an alternative method of determining the total gas mass is needed. 

The second commonly used method to determine disk mass is 
to model the long wavelength dust emission for an assumed dust opacity and dust temperature and then convert to a gas mass assuming a gas-to-dust ratio, typically taken to be the ISM value of 100 \citep{Williams11}. 
If the gas-to-dust ratio differs from that in the ISM this method becomes less reliable.  Such would be the case if a significant fraction of the dust has been incorporated into large, $>$ cm-sized grains or planetesimals, which do not contribute to the observed sub-mm continuum.

Uncertainty in the gas-to-dust ratio also plays into constraints on the gas surface density profile. Often the gas surface density is taken to follow the dust surface density, itself derived from resolved continuum observations or SED fitting \citep{Calvet02,Guilloteau11}.
Assuming the same surface density profile for gas and dust is likely insufficient, particularly in systems where the gas emission is far more extended than emission from large, millimeter-sized grains \citep{Andrews12}. 
Though there have been efforts to constrain the surface density based on spectral line observations these efforts often require comparison to existing models due to limits on the spatial resolution of the data as well as an understanding of the particular species abundance relative to the total gas mass, which is complicated by chemistry \citep{vanZadelhoff01}.
A robust surface density measurement thus requires observations of a species whose abundance relative to \hh\ is well known.

Recently, \citet{Bergin13} detected the HD J=1-0 ($E_u = 128$ K) line towards the 3-10~Myr old transition disk TW Hya \citep{Barrado06,Vacca11} using the Herschel Space Observatory. This spatially and spectrally unresolved detection provides a gas mass tracer more closely related to \hh\ than either CO or dust.
Previous gas mass estimates for TW Hya range from $5\ee{-4}$ to $0.06$~\msun \citep{Calvet02,Thi10,Gorti11}.
Using the HD detection \citet{Bergin13} find the total gas mass in the TW Hya disk to be $> 0.05~\msun$, significantly larger than most disk mass estimates derived from CO emission. 

With an independent method of deriving the \hh\ mass, the CO abundance relative to \hh\ was measured to be $X\left(\mathrm{CO}\right) = \left( 0.1-3\right) \ee{-5}$ using partially spatially resolved observations of \cooo~2-1, significantly below the canonical ISM value of $\eten{-4}$ \citep{Favre13}. 
Comparison of the azimuthally averaged CO surface density calculated from resolved observations of \cooo\ 3-2 with resolved dust continuum shows that this depletion extends inward at least as far as 10 AU \citep{Nomura15}.
The recent detection of \ion{C}{1} in TW~Hya confirms that \ion{C}{1} is also under-abundant by roughly a factor of 100 in the outer disk \citep{Kama16}. 
This evidence, along with modeling of TW Hya tailored to match a suite of observations, further suggests a global depletion of the volatile gas phase carbon in this system, rather than a low CO abundance due to in situ chemical processes such as photodissociation or freeze-out \citep{Cleeves15,Du15}.

Direct measurements of the CO abundance relative to \hh\ in TW Hya hinge on the calculated gas mass based on the detection of HD. Both the derived gas mass and CO abundance are highly dependent on the assumed gas thermal structure, with the fractional CO abundance varying by a factor of 30 for assumed gas temperatures from 20-60 K \citep{Favre13}. Knowledge of the thermal structure in the HD emitting layers is essential to better constrain both the total gas mass, the gas surface density profile, and the CO abundance.
Previous spectrally resolved observations of low- and high-J CO in disks have been used to measure the vertical temperature structure and, by comparing with models, constrain the radial structure \citep{vanZadelhoff01,Dartois03,Fedele13}. The spatially and spectrally resolved observations of TW Hya presented here allow us to determine the radial temperature structure in this nearly face-on disk ($i\sim7\degree$ \citet{Qi04}), directly from the data. 

The thermal structure in the disk also impacts the chemical structure. 
Snowlines in disks for a given species occur where the rate of adsorption onto a grain surface equals the rate of desorption. The exact location, both radially and vertically, depends directly on the thermal structure in the disk, with the snowline for a given species existing at larger radii for warmer disks.
Snowlines can be observed directly, using emission from optically thin isotopologues such as \cooo\ 2-1 in the Herbig disk HD 163296,
or indirectly using tracers such as \nnhp\ in TW Hya \citep{Qi13b,Qi15}.
 Understanding the radial thermal structure in protoplanetary disks is vital to our understanding  of the growing number of observations with resolved molecular emission structure. 

In this work, we present resolved observations of the TW Hya disk ($d=54\pm6$ pc) in \coo\ 3-2, \cooo\ 3-2, \coo\ 6-5, and \cooo\ 6-5 line emission carried out with the Atacama Large Millimeter/submillimeter Array (ALMA). Using these observations we obtain a high resolution estimate of the radial CO abundance structure in a protoplanetary disk in addition to detecting the surface CO snowline.
\S \ref{observations} details the observations and data reduction process, while \S \ref{analysis} briefly summarizes the observational results and details how we derive the radial gas temperature structure. This is then used to calculate the \hh\ surface density, the radial CO abundance profile, and estimate the location of the midplane CO snowline.
We discuss the implications of these findings in \S \ref{discussion} as well as the possible causes of the emission structure seen in the data. Finally, our results are summarized in \S \ref{summary}.

\begin{figure*}
\setlength{\intextsep}{0pt}
    \includegraphics[width=1.0\textwidth]{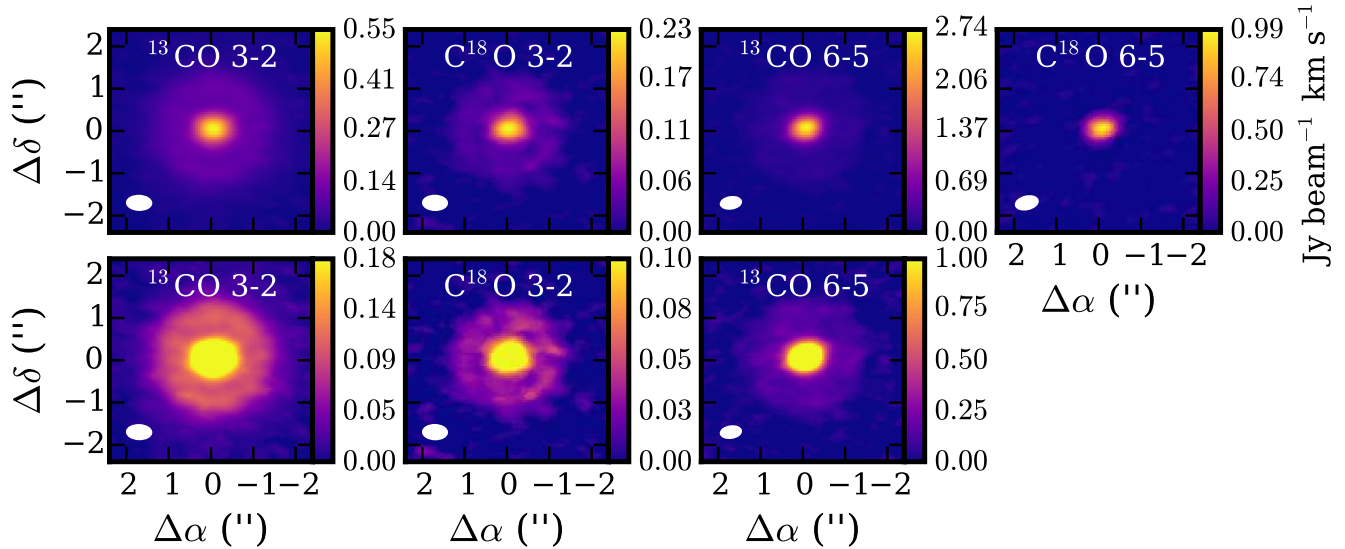}
  \caption{Top: Integrated emission maps of \coo\ 3-2 (peak flux 0.53 Jy beam$^{-1}$ km s$^{-1}$) , \cooo\ 3-2 (0.22 Jy beam$^{-1}$ km s$^{-1}$), \coo\ 6-5 (2.16 Jy beam$^{-1}$ km s$^{-1}$), and \cooo\ 6-5 (1.04 Jy beam$^{-1}$ km s$^{-1}$). Bottom: The same integrated emission maps rescaled to pull out on the extended emission beyond 0\farcs5. \cooo\ 6-5 is not detected beyond 0\farcs37.}
  \label{codata}
\end{figure*}

\section{Observations and Data Reduction}\label{observations}
ALMA observations of TW Hya were obtained in Band~9 on March 12, 2014 with 27 antennas and in Band~7 on May 14, 2015 with 37 antennas. The baseline coverage was 15-414 m for the Band~9 observations and 21-545 m for those in Band~7. 

Observations in both bands utilized 4 spectral windows (SPWs). For Band 7 the spectral resolution for SPW1 and SPW2 was 122.070 kHz with a total bandwidth of 468.75 MHz. These windows contained the \coo\ 3-2 line and the \cooo\ line, respectively. SPW3 covered the \coo\ 3-2 and \cooo\ 3-2 lines with 488.281 kHz spectral resolution and 1.875 GHz bandwidth. The final spectral window had a resolution of 15.625 MHz and a bandwidth of 2.0 GHz.
For Band 9 the spectral resolution for SPW1 and SPW2, containing the \coo\ 6-5 and \cooo\ 6-5 lines respectively, was 244.141 kHz and the total bandwidth was 937.5 MHz.  The spectral resolution and bandwidth for SPW3 and SPW4 were the same as for Band 7.

\begin{figure}[h!]
\setlength{\intextsep}{0pt}
    \includegraphics[width=0.45\textwidth]{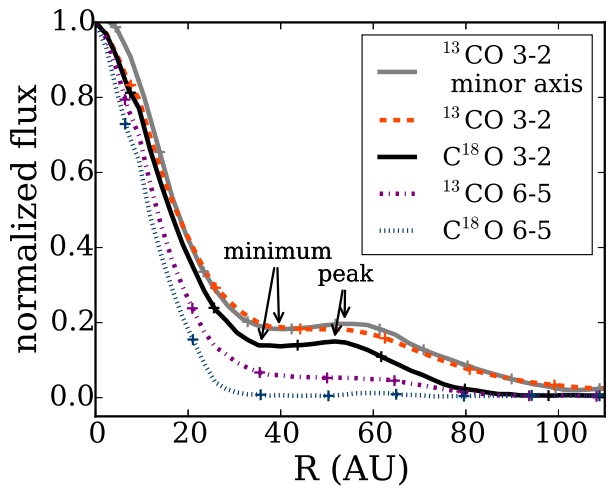}
  \caption{Deprojected azimuthally averaged \coo\ 3-2, \cooo\ 3-2, \coo\ 6-5, and \cooo\ 6-5 integrated emission normalized to the peak. Crosses are points with half beam separation. The 3-2 emission minima and secondary peaks are highlighted.}
  \label{avgI}
\end{figure}

On both dates Titan was used for amplitude and flux calibration, while phase calibration and bandpass calibration were carried out on J1037-2934 and J1256-057 respectively.
Initial data reduction was carried out by ALMA/NAASC staff using standard procedures. In addition, phase and amplitude self-calibration were carried out for the spectral windows containing the science targets using CASA 4.6.12. For the Band 9 observations, self-calibration was carried out separately for each SPW. Continuum subtraction was employed for each spectral window using the line free channels for all observations.

A CLEAN mask was manually generated individually for each spectral window with line emission. Briggs weighting with the robustness parameter set to 0.5 was used for the \coo\ and \cooo\ 3-2 transitions; natural weighting was used for the \coo\ and \cooo\ 6-5 emission. The restoring beam for the Band 7 observations had FWHM dimensions of $\sim 0\farcs5 \times 0\farcs3$ (P.A. 88\degree), while those for the \coo\ and \cooo\ 6-5 maps were
$\sim 0\farcs4 \times 0\farcs2$ (P.A. -85\degree) and $\sim 0\farcs5 \times 0\farcs3$ (P.A. -77\degree) respectively. The slightly different beams for the Band 9 observations are the result of the self-calibrations preformed separately for each SPW. The RMS of the final CLEANed images in a 0.1 km s$^{-1}$ channel are 9.2 mJy beam$^{-1}$ for \coo\ 3-2, 12 mJy beam$^{-1}$ for \cooo\ 3-2, 56 mJy beam$^{-1}$ for \coo\ 6-5, and 77 mJy beam$^{-1}$ for \cooo\ 6-5. 
Integrated emission maps were made by summing the emission above 2$\sigma$ from each channel. 

\section{Data Analysis}\label{analysis}
Figure~\ref{codata} shows the integrated emission maps for \coo\ 3-2, \cooo\ 3-2, \coo\ 6-5, and \cooo\ 6-5 in TW Hya while
Figure~\ref{avgI} shows the de-projected azimuthally averaged emission profiles. The \coo\ 3-2, \cooo\ 3-2, and \coo\ 6-5 lines all show a plateau of weak extended emission in addition to the bright, centrally peaked emission. The \coo\ 6-5 and \cooo\ 3-2 emission extends to 1\farcs30 while that for \coo\ 3-2 extends to 1\farcs87. The \cooo\ 6-5 emission extends to only 0\farcs37. 
Additionally, the \coo\ 3-2 and \cooo\ 3-2 transitions show a flux decrease near $\sim0\farcs73$ (about 40~AU) and $\sim0\farcs66$ (37~AU) respectively with the outer ring of emission peaking at $\sim1\farcs0$ (54~AU) and $\sim0\farcs95$ (51~AU). This feature is not seen in the 6-5 data, though this may be due to insufficient sensitivity. The radius of the emission minimum appears to vary with azimuth, most likely a result of the varying resolution along different axes due to the ellipsoidal beam. This results in the feature being smoothed out in the azimuthally averaged \coo\ 3-2 emission profile but clearly seen in the average profile along the minor axis (Figure~\ref{avgI}). The average radius of the gap and the ring are $0\farcs70$ (38~AU) and $0\farcs97$ (53~AU) respectivley, with the uncertainty due to the resolution of the observations much greater than the difference between the two lines. There have been many previous detections of molecular emission rings, including hydrocarbon features \citep{Qi13b,Kastner15,Oberg15}. Our data reveal ring structure in CO, a fundamental tracer of the total gas phase carbon. 

\subsection{Gas Temperature }
The optical depth of the \coo\ lines are determined by taking the ratio of the \coo\ and \cooo\ emission in each transition on a pixel-by-pixel basis.
The ratio of the observed \coo\ and \cooo\ intensity for a given transition can be directly related to the excitation temperature and the optical depth:
\begin{equation}
\frac{T_{B}\left(^{13}\mathrm{CO} \right)}{T_{B}\left(\mathrm{C^{18}O} \right)} = \frac{T_{ex,^{13}\mathrm{CO}}\left(1-\mathrm{e}^{-\tau_{^{13}\mathrm{CO} }}\right)}{T_{ex,\mathrm{C^{18}O}}\left(1-\mathrm{e}^{-\tau_{\mathrm{C^{18}O}}}\right)}
\end{equation}
$T_{ex,^{13}\mathrm{CO}}/T_{ex,\mathrm{C^{18}O}}$ is expected to be close to unity, effectively canceling. The ratio of the optical depths is equal to the ratio of their abundances, assumed to be \coo/\cooo\ = 8 based on ISM abundances \citep{Wilson99}, allowing us to solve for the optical depth.\footnote{This method for determining optical depth only works when one of the lines is optically thick.}
We find that \cooo\ is optically thin in both transitions throughout the disk, $\tau=0.7-0.3$ for the 3-2 and $\tau=0.5-0.4$ for the 6-5, while \coo\ 3-2 is optically thick everywhere, $\tau=4.5-1.9$. \coo\ 6-5 is optically thick inside 0\farcs37, in the range $\tau=3.6-1.9$, and assumed to be optically thin beyond 0\farcs37, where the lack of \cooo\ 6-5 emission
prevents us from calculating the optical depth.
It is possible that isotopologue-specific photodissociation has enhanced the \coo/\cooo\ ratio. Such an enhancement would decrease the calculated optical depth for all transitions, meaning that the values used here are upper limits. The effects of isotopologue-specific photodissociation are explored further in \S \ref{discussion}.

The optically thick \coo\ 3-2 and 6-5 emission is used to measure the kinetic gas temperature in the disk. 
Optically thick spectral lines have long been used as temperature probes \citep{Penzias72}.
The observed intensity can be related to the kinetic gas temperature, $T_{K}$, assuming LTE:
\begin{equation}
T_{B} = \frac{h \nu \left(1-\mathrm{e}^{-\tau_{\nu}} \right)}{k \exp{\left(h \nu/k T_{K}-1\right)}}
\end{equation}
assuming the emission fills the beam, as is reasonable for resolved emission.
For our data the temperature was calculated for each pixel before taking the azimuthal average (cf. Figure~3a).

The apparent decrease in the temperature profile inside of 10 AU is due to the large velocity spread in the inner disk. While the integrated emission is centrally peaked, the peak channel flux per pixel, which is used to calculate the gas temperature, is maximized near 10 AU. Thus, the decreasing temperature profile in the inner disk does not reflect an actual decrease in temperature, though the temperature is well constrained from 10-60 AU due to the emergence of single peaked profiles.

The temperature profiles derived from \coo\ 3-2 and \coo\ 6-5 (Figure~3a) show similar structure but differ in absolute value with the 6-5 finding higher temperatures relative to the 3-2. This is well known in the sense that the 6-5 emission has excitation characteristics that lead to the transition becoming optically thick at higher altitudes than the 3-2; these higher layers are closer to the heated surface and are hence warmer. This can been seen quite readily in the detialed modeling of \citet{Bruderer12}.

Similar to the emission profiles, the temperature profile plateaus in the outer disk. This is a direct consequence of the temperature tracers used. In the warm inner disk CO is able to exist in the gas phase throughout the disk. 
The emission originates from a wide range of heights and, thus, temperatures.
In the outer disk CO is emitting from vertical region in the disk often referred to as the warm molecular layer and probes a much narrow range in height \citep{Aikawa02}.
The flatness of the temperature profile indicates that most of the emission we detect originates from the layers of the disk just above the freeze-out temperature.
\textit{Thus, we have effectively detected the surface snowline in the cold outer disk.}
For subsequent calculations we use the average of the two temperatures where available. In the outer disk where \coo\ 6-5 is optically thin we rely on the temperature derived from the 3-2 observations. The derived temperature profile is shown in Figure~3a.

%\begin{figure*}
%\setlength{\intextsep}{0pt}
%\subfloat{\label{temp}\includegraphics[width=0.45\textwidth]{f3a.eps}}
%\subfloat{\label{sigmagas}\includegraphics[width=0.45\textwidth]{f3b.eps}}
%\\
%\subfloat{\label{SigmaCO}\includegraphics[width=0.45\textwidth]{f3c.eps}}
%\subfloat{\label{cotohh}\includegraphics[width=0.45\textwidth]{f3d.eps}}
%\label{megaplot}
%\caption{a) Average radial temperature profile derived from the optically thick \coo\ emission. b) Surface density of the warm gas as traced by HD. c) Derived surface density of CO using the average temperature profile. The shaded region indicates the limit assuming the temperatures measured from \coo\ 3-2 and \coo\ 6-5. d) Azimuthally averaged CO abundance relative to \hh. Crosses are points with half beam separation.}
%\end{figure*}

\begin{figure*}[t!]
\setlength{\intextsep}{0pt}
\begin{subfigure}
\centering
\includegraphics[width=0.45\textwidth]{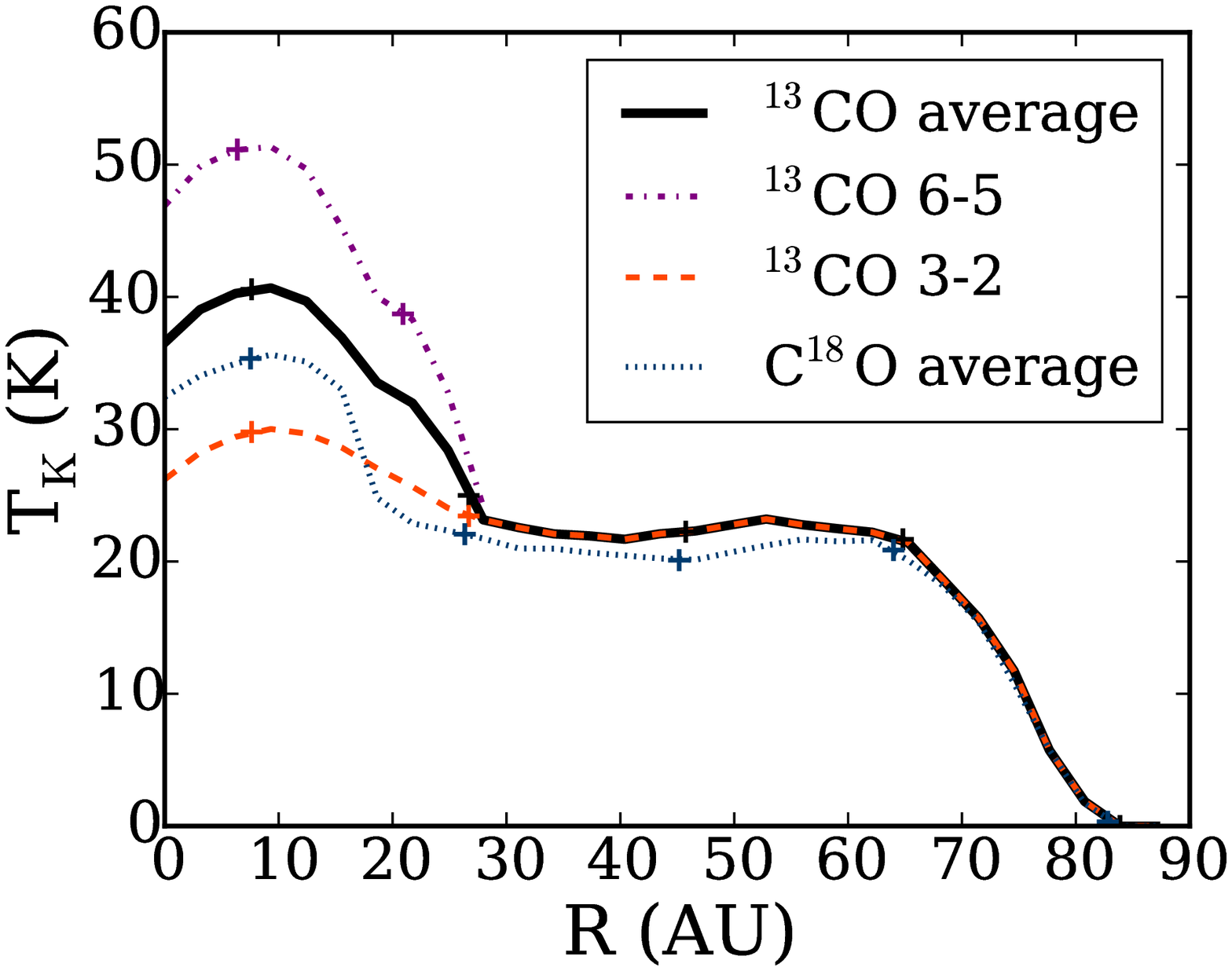}
\end{subfigure}
\begin{subfigure}
\centering
\includegraphics[width=0.45\textwidth]{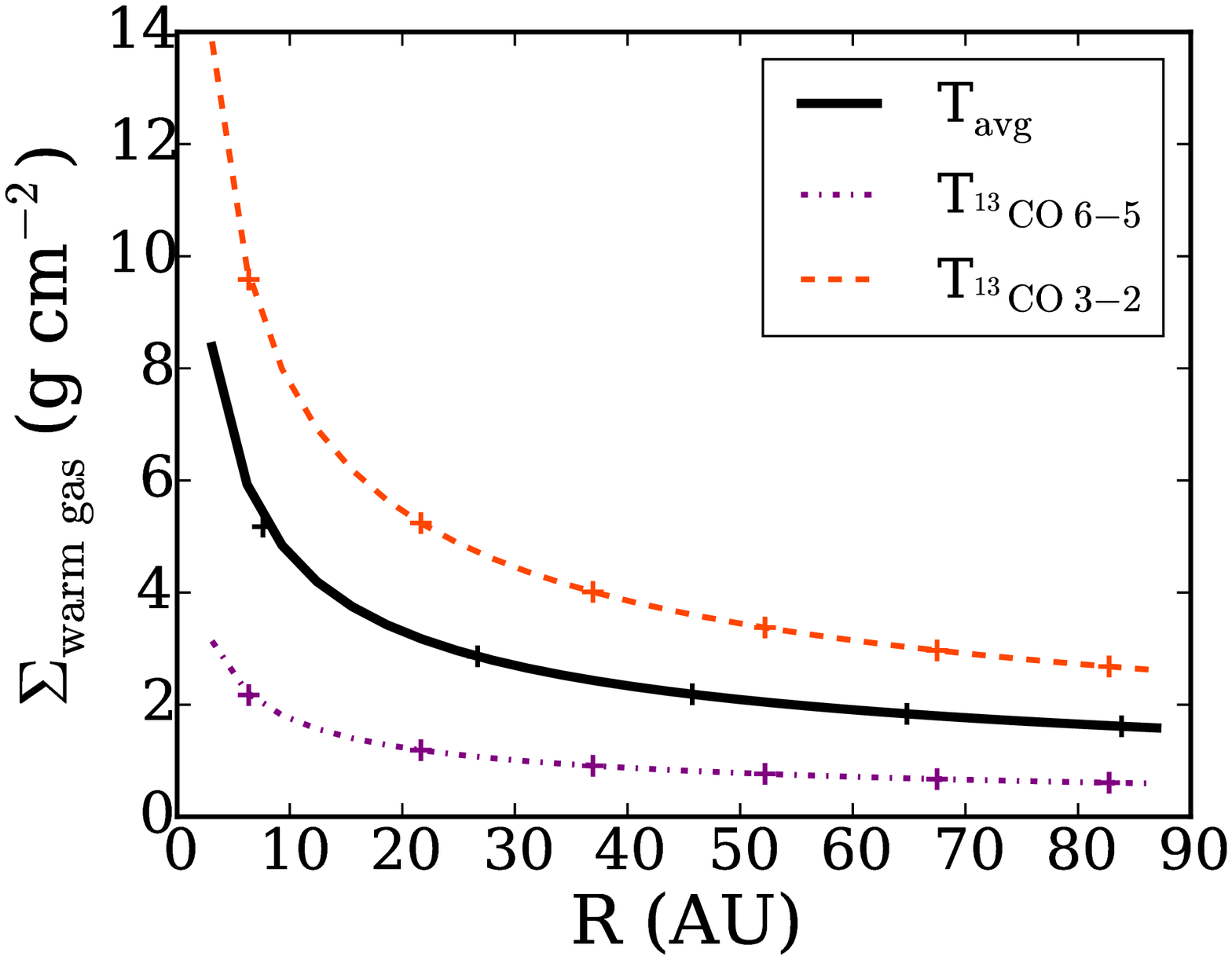}
\end{subfigure}
\\
\begin{subfigure}
\centering
\includegraphics[width=0.45\textwidth]{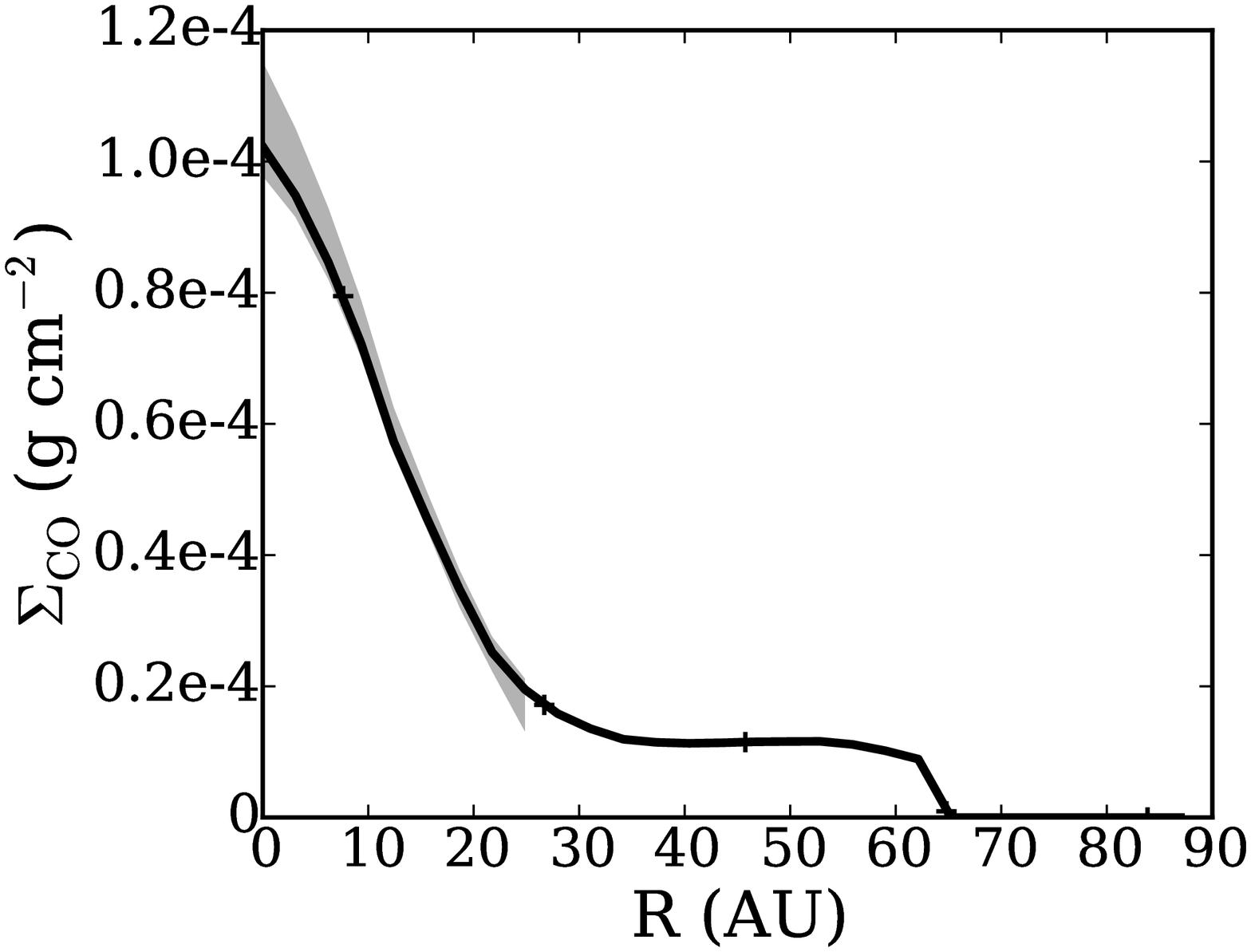}
\end{subfigure}
\begin{subfigure}
\centering
\includegraphics[width=0.45\textwidth]{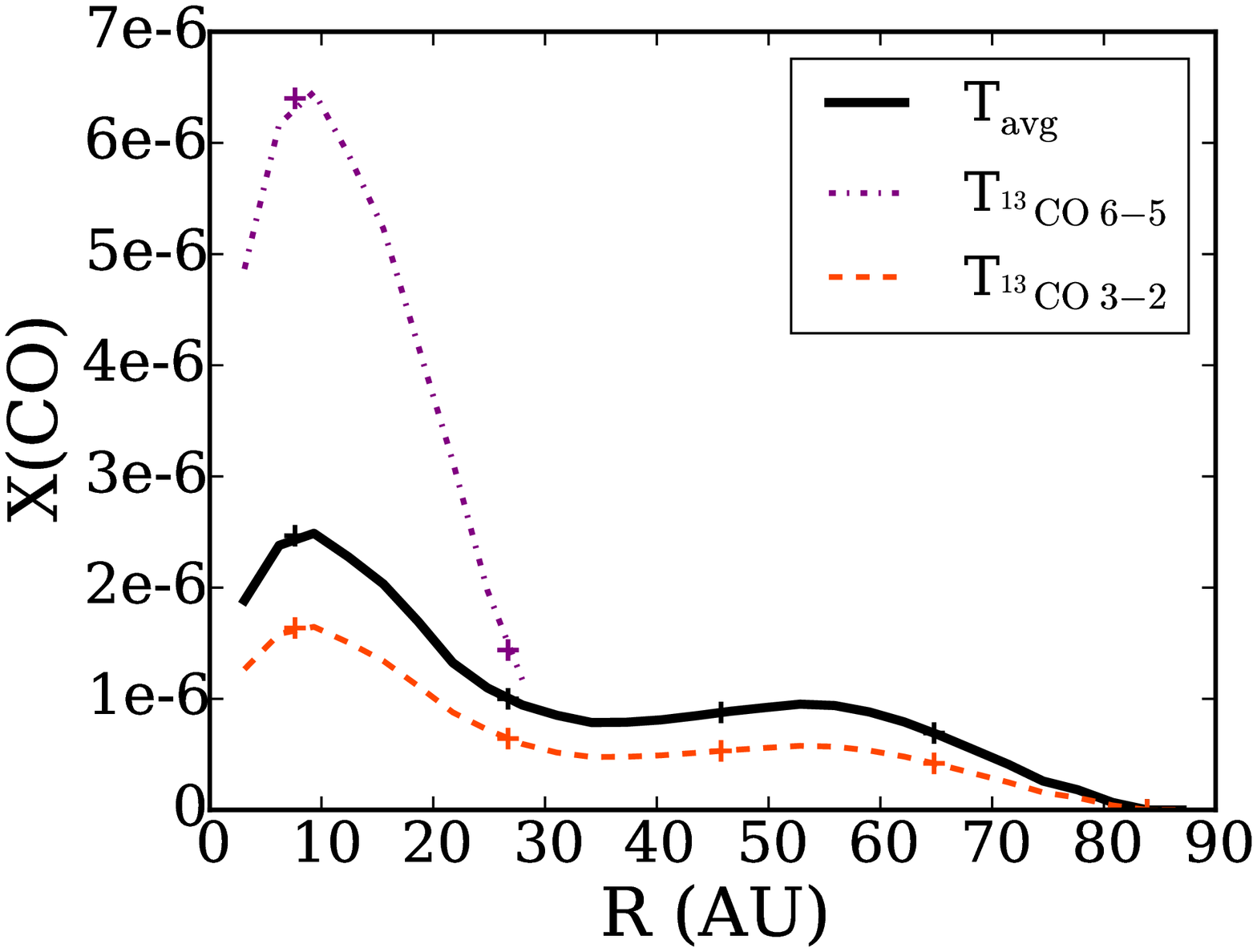}
\end{subfigure}
\label{megaplot}
\caption{a) Average radial temperature profile derived from the optically thick \coo\ emission. b) Surface density of the warm gas as traced by HD. c) Derived surface density of CO using the average temperature profile. The shaded region indicates the limit assuming the temperatures measured from \coo\ 3-2 and \coo\ 6-5. d) Azimuthally averaged CO abundance relative to \hh. Crosses are points with half beam separation.}
\end{figure*}

\subsection{The H$_2$ Surface Density Distribution}
The temperature derived above, along with the previous detection of HD 1-0 towards TW Hya, allow us to calculate the total warm gas surface density without relying on an assumed CO abundance.
While there have been previous mass measurements based on the HD detection, we can now calculate the surface density using a measured gas temperature.
In addition to calculating the gas surface density using our average \coo\ temperature profile as our reported value we also calculate limits on the surface density using the temperature profile from only the \coo\ 6-5 data and only the \coo\ 3-2 data. 

The \coo\ 6-5 transition has an upper state energy ($E_{u}/k= 111.05$) similar to that of HD~1-0 ($E_{u}/k= 128.49$) and provides an upper limit for the temperature of the HD emitting gas. \coo\ 3-2 is likely the best match to the thermal conditions of the disk as the \co\ 3-2 line is clearly stronger than the \co\ 2-1 line based on previous ALMA observations of this system \citep{Rosenfeld12}.
There are likely regions of the disk cooler than those traced by the \coo\ 3-2 line. However, because it emits at temperatures above 20~K, HD only traces the warm gas. Since we lack information on the full thermal structure of the disk, the values reported below are a lower limit on the total gas mass.
We assume that the warm, $>$20~K, gas in which HD is emitting follows the surface density profile of the small ($r<100\ \micron$) dust as fit by \citet{Menu14} i.e., $\Sigma \propto R^{-1/2}$. The impact of this assumption is discussed in \S \ref{altsurface}. 

Using our measured average \coo\ temperature profile we calculate the strength of the HD 1-0 emission in radial bins, assuming \hh\ has the same surface density as the small dust grains and an abundance ratio of HD/\hh$ = 3\ee{-5}$ \citep{Linsky98}.
We then integrate the emission over the disk, taking the ratio of the unresolved HD detection and the calculated integrated emission. This yields a scaling factor of 73 for the average temperature profile. The surface density profile is then uniformly scaled by this factor, such that the total calculated emission agrees with observations (Figure~3b).
The final surface density profile for the warm gas is
\begin{equation}
\Sigma_{warm\,gas} = 4.7^{+3.0} _{-2.9}\mathrm{\,g\, cm^{-2}}\left(\frac{R}{10\,\mathrm{AU}}\right)^{-1/2} 
\end{equation}
in the range $ 3.1\,\mathrm{AU}\leq R \leq 61.7\, \mathrm{AU}$ and zero elsewhere. The uncertainties indicate how the derived surface density changes if we use the temperature from \coo\ 3-2 (mass upper limit) and the temperature from \coo\ 6-5 (lower limit). The fall off is assumed to go as $R^{-1/2}$.
In the inner disk, where the temperature exceeds 20 K at all heights, HD is sensitive to the total gas column except that near the midplane where $\tau_{112\mathrm{\mu m}} > 1$, masking some fraction of the emission.
In the outer disk where the midplane is cooler than 20 K, HD is a less sensitive probe of the total gas mass.
Thus, the surface density we derive should be considered a lower limit, constraining the gas surface density in TW~Hya for the first time.

We find that the total warm gas mass inside of 61.7 AU is  $5.6\ee{-3}~\msun$ assuming our average temperature profile.
Assuming that all of the HD 1-0 emission emits at the measured \coo\ 6-5 temperature reduces the warm gas mass to $2.1\ee{-3}~\msun$ while assuming the \coo\ 3-2 temperature increases the mass to $9.3\ee{-3}~\msun$ since at cooler temperatures more gas is needed to produce the same emission.
Previous modeling of TW Hya indicates that approximately 10-20\% of the total gas mass is above 20 K \citep{Andrews12,Bergin13}. As such the total disk gas mass based on our calculations is of order $5.6\ee{-2} \msun$, consistent with the lower limit of $>0.05 \msun$ established by \citet{Bergin13}. 

\subsection{CO Surface Density \& Abundance}
With a measured gas temperature in hand we use the optically thin \cooo\ 3-2 and 6-5 emission, as well as the optically thin \coo\ 6-5 emission in the outer disk, to calculate $N_{J=6}$ and $N_{J=3}$ for CO:
\begin{equation}
N_{u} = \frac{8 \pi k \nu_{0}^{2} T_{B}}{h c^{3} A_{ul}}
\end{equation}
where $N_{u}$ is the upper state column density, $\nu_{0}^{2}$ is the rest frequency of the transition, $T_{B}$ is the peak brightness in Kelvin, and $A_{ul}$ is the Einstein A coefficient for the transition. 

After converting to a \co\ abundance, we correct for the fractional population not in the J=6 or J=3 states using our average \coo\ temperature profile:
\begin{equation}
N = \frac{N_{6} + N_{3}}{f_{6}+f_{3}}
\end{equation}
where the fractional upper state population, $f$ is:
\begin{equation}
f_{u} = \frac{g_{u}}{Q \mathrm{e}^{\Delta E/kT}}
\end{equation}
and the partition function, $Q$, approximated for a linear rotator is:
\begin{equation}
Q = \frac{k T}{h B_{0}} + \frac{1}{3}.
\end{equation}
We are then able to calculate the surface density of CO in TW Hya (Figure~3c). Our CO surface density is similar to the \citet{Nomura15} surface density derived solely from \cooo\ 3-2, though our profile peaks at a lower value and displays a flatter slope between 40-60 AU. These discrepancies are primarily due to the different temperature profiles used.

Using the derived \hh\ surface density in conjunction with the measured CO surface density we map the CO abundance relative to \hh\ in TW Hya as a function of radius (Figure~3d). The improved spatial resolution of our observations show that CO is indeed universally depleted; everywhere we detect CO emission $X(\mathrm{CO})$  is of order $\eten{-6}$, consistent with the previously found global CO abundance \citep{Bergin13,Favre13,Cleeves15}. Though some CO returns to the gas phase inside the snowline, $X(\mathrm{CO})$  never rises above $2.5\ee{-6}$ assuming the average \coo\ temperature profile. Clearly there is a significant amount of gas phase carbon missing from the observable TW Hya disk.

\section{Discussion}\label{discussion}
\subsection{The H$_2$ Surface Density}\label{altsurface}
\begin{figure} [h!]
\setlength{\intextsep}{0pt}
    \includegraphics[width=0.45\textwidth]{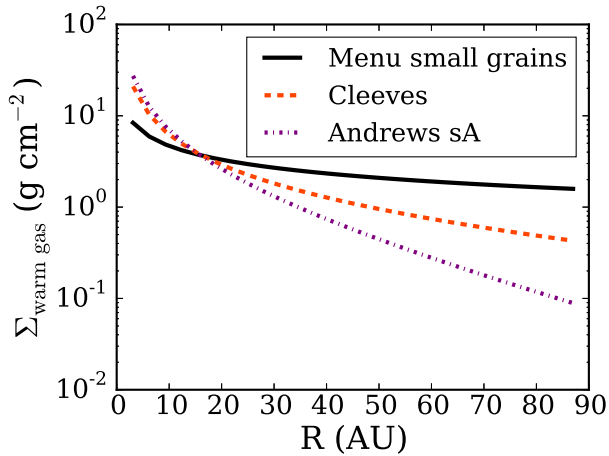}
  \caption{Comparison of different radial models for the total warm gas surface density as traced by HD.}
  \label{allsigma}
\end{figure}

One of the major assumptions in the above analysis is that the \hh\ surface density follows that of the small dust grains from the modeling of \citet{Menu14}.
Here we explore the effect of alternative surface densities. 
In particular, we consider
the best fit model for the observed \co\ 3-2 emission from the modeling efforts of \citet{Andrews12}, their model sA, as well as the dust surface density profile of \citet{Cleeves15}. The derived surface density of the warm gas assuming each of these radial profiles is shown in Figure~\ref{allsigma}. 
Because it is normalized to match the HD emission, we find that the surface density of the warm gas varies by less than a factor of three for $R=1-40$~AU. 
As such, the low CO abundance cannot be explained by the uncertainty of the gas surface density profile.

\subsection{The Missing Carbon}
Models demonstrate that self-shielding of CO is capable of modifying the gas phase CO isotopologue ratio in both the ISM and protoplanetary disks \citep{vanDishoeck88,Visser09}.
Recent work by \citet{Miotello14} shows that self-shielding can raise the \co/\cooo\ ratio by up to an order of magnitude for vertical layers in which the \co\ has become optically thick in the UV while \cooo\ is still exposed to photo-dissociating radiation. Thus, any observations using \cooo\ as a tracer of the total CO abundance potentially under-predict the total abundance by an order of magnitude. This would partially, but not completely, explain our low CO abundance. 

To check whether self-shielding is important for our observations we compare the \coo\ and \cooo\ abundances in the outer disk as traced by \coo\ and \cooo\ 3-2. We find a \coo/\cooo\ ratio in the range 10-12 in the outer disk, compared to the ISM value of 8. 
If we assume that \coo\ is completely self-shielded, and thus that the \co/\coo\ ratio is similar to the ISM, then \co/\cooo\ $\sim 690-830$, a factor of 1.2-1.5 greater than in the ISM. 
This is comparable to the disk averaged \cooo\ depletion in a \eten{-2} \msun\ disk as found by \citet{Miotello14}.
However, assuming \co/\cooo\ $=830$ only increases the $X(\mathrm{CO})$  maximum to $3.2\ee{-6}$.

Similar to the CO isotopologues, \hh\ self-shields before HD, meaning there is a region of the disk where the HD/\hh\ ratio is smaller than what is assumed here. \citet{Bergin14} investigate this possibility using disk chemical models and find that the region in which HD has self-shielded but \cooo\ has not accounts for $<1\%$ of the total mass. Thus, self-shielding of HD cannot explain the low CO abundance.

Carbon must therefore be removed from gas phase CO. Two plausible routes are chemical reprocessing and freeze-out onto grains. In regions of the disk exposed to X-rays from the central star He$^+$ can react with CO to create C$^+$, a fraction of which is incorporated into CO$_{2}$ and hydrocarbons \citep{Aikawa97,Bergin14,Reboussin15}. Often, these species are able to freeze-out onto grains at temperatures where CO primarily resides in the gas phase, effectively removing carbon from the gas phase chemistry.

In addition to chemical processing, it is possible that vertical mixing is able to deliver gas phase CO or CO$_{2}$ to the midplane. There it freezes out onto grains too large to be lofted to warmer layers. The freeze-out of volatiles also explains the low oxygen abundance in TW Hya \citep{Du15}. 
There is a clear decrease in the emission and fractional CO abundance around 30 AU, which is associated with the CO snowline (Figures~\ref{codata}~\&~3d). However, even inside 30~AU very little carbon is returning to the gas.
It is possible that the CO abundance is greater at radii much smaller than our resolution of 13.5 AU. However, $\sim\eten{-7}$~\msun\ of gas phase CO would need to reside in the inner few AU to fully explain the overall depletion, which is unlikely.

This depletion of CO is also seen in at least one other tracer of gas phase carbon. For \ion{C}{1}, \citet{Kama16} find a factor of 100 reduced abundance with respect to the ISM. Another tracer of carbon would be C$^+$. Using current upper limits \citep{Thi10}, a temperature of 40 K (Figure~3a), and HD to trace \hh\ (mirroring the analysis of \citet{Favre13}) we find that the abundance limit for C$^+$ is $<7.5\ee{-4}$, well above the ISM abundance of carbon \citep{Langer14}. However, the depletion of both CO and \ion{C}{1} suggests that carbon is largely not returning to the gas phase, even in layers above the nominal CO freeze-out temperature.
If this is in fact due to freeze-out the volatile carbon (e.g. CO, CO$_2$, or simple hydrocarbons) would need to be locked inside bodies large enough to avoid destruction via evaporation at small radii.

\subsection{The CO snowline and Outer Ring}
The moment zero maps in Figure~\ref{avgI} are characterized by bright centrally peaked emission and a plateau of weaker extended emission. The notable exception is the \cooo\ 6-5 map, which shows only the central emission. 
The transition to the plateau of emission occurs between 20-35 AU, roughly the radius of the CO snowline at $R\sim 30$~AU as traced by \nnhp\ 4-3 \citep{Qi13b}. 

Determining the precise location of the snowline is difficult due to the radial and vertical structure in the disk as well as limitations imposed by the spatial resolution of the observations. We outline two methods for characterizing the snowline with the data in hand. The first is to read the location directly from the surface density profile. The second is to calculate the expected radius for a given gas and temperature structure. 

We find the surface snowline to be $R\sim30$~AU based on the CO surface density profile (Figure~3c). 
Beyond the CO snowline our observations are only able to probe gas above the CO freeze-out temperature and will be biased towards the vertical layers with the highest CO gas density. This occurs right above the CO freeze-out surface, i.e., at radially increasing heights in the disk where CO gas freezes onto dust grains. The average \cooo\ temperature profile indicates that the CO sublimation temperature in this system is slightly less than 21 K (Figure~3a). This results in an observed surface density profile which is roughly constant outside of $R\sim30$~AU. 
It should be noted that this is not the midplane snowline radius. 
Rather it is the snowline at the vertical height in the disk traced by CO isotopologues in the J=3 and J=6 states, much nearer the surface than the midplane \citep[e.g.][]{Dent13}. In a passively heated disk this radius will be greater than the radius of the midplane CO snowline.

The snowline is a chemical/physical transition in the disk, occurring where the rate of deposition onto and sublimation off of a grain surface are equal.
Combining knowledge of the CO surface density and binding energy derived directly from our observations with existing models of the disk structure in TW Hya we are able to estimate the location of the midplane CO snowline.
The scale height at each radius is calculated using our average temperature profile derived from \cooo, which probes nearer the midplane than the optically thick \coo, and assuming a central stellar mass of 0.8 \msun \citep{Wichmann98}:
\begin{equation}
H = \sqrt{\frac{kT_{K} R^{3}}{2.3 m_H G M_{*}}}.
\end{equation}
Using our measured CO surface density, which allows us to account for the observed CO depletion, the number density of CO molecules is then:
\begin{equation}
n_{CO}\left(R,Z\right) = \frac{\Sigma_{CO}}{m_{CO}\,\sqrt{2 \pi} H} \exp{\left[-\frac{1}{2}\left(\frac{Z}{H}\right)^{2}\right]}.
\end{equation}
We solve for the temperature at which the adsorption and desorption fluxes are equal, assuming the gas and dust temperatures are the same:
\begin{equation}
T_{K} = \frac{E_{B}}{k} \ln{\left[\frac{4Nf \nu}{n_{\mathrm{CO}} v}\right]},
\end{equation}
where $f\sim 1$ is the fraction of absorption sites occupied by CO, $\nu$ is the vibrational frequency of CO in the surface potential well, $v$ is the thermal speed of CO,
$E_{B}$ is the binding energy for CO on an ice coated surface, and $N = \eten{15}$ is the number of absorption sites per cm$^2$, as is appropriate if 10 mm$^2$ of surface area per cm$^2$ is available for freeze-out, assuming each molecule occupies 1 \AA$^2$ on the grain surface \citep{Hollenbach09}. 
A freeze-out temperature of 21 K, the temperature at which the average \cooo\ temperature profile plateaus, suggests $E_{B}/k\sim960$ K.
This derived binding energy is consistent with laboratory measurements of CO binding to a primarily CO ice surface, perhaps with some contamination from \water\ and CO$_2$ ice \citep{Collings03,Oberg05,Cleeves14}.

Our derived temperature profile is not a good probe of the midplane temperature in the disk, being more sensitive to the warmer vertical layers in the disk, and provides only an upper estimate for the midplane temperature. To better constrain the radius of the midplane snowline we use the midplane gas temperature from the TW Hya model of \citet{Cleeves15}. Substituting these temperatures into Equation 10 and using the binding energies derived above we calculate a midplane CO snowline radius in the range $R = 17-23$ AU. We stress that this result is model dependent. Assuming a different midplane temperature or density structure would shift the calculated midplane snowline radius.

In addition to the drop in emission at the surface CO snowline both the \coo\ and \cooo\ 3-2 integrated emission maps show a deficit of emission centered at $R\sim38$~AU $(0\farcs70) $ with the emission beyond this minimum peaking at $R\sim53$~AU $(0\farcs97)$ (Figure~\ref{avgI}). The minimum in the \cooo\ 3-2 emission is 31 mJy beam$^{-1}$ km s$^{-1}$, 2.6 times the RMS, while the secondary peak is 34 mJy beam$^{-1}$ km s$^{-1}$. The \coo\ 3-2 minimum is 10.7 times the rms, 99 mJy beam$^{-1}$ km s$^{-1}$, with the secondary peak at 106 mJy beam$^{-1}$ km s$^{-1}$.
This feature can also be seen in the \coo\ and \cooo\ emission maps of \citet{Nomura15} at $>5\sigma$.

The explanations for this ring fall into two categories: processes that result in additional depletion of CO near $R\sim36$~AU and processes that result in the return of gas phase CO in the outer disk. 
Rapid grain growth near the CO snowline could trap CO ices beneath the surface of grains \citep[e.g.][]{Ros13}, preventing them from returning to the gas phase while CO ice in regions without rapid grain growth will remain on the grain surface, subject to photodesorption. Indeed, there is a bump in the dust emission profile near this radius, consistent with such grain growth \citep{Nomura15,Zhang16}.

Alternatively, the CO could be returning to the gas at large radii due to changing physical conditions.
The outer ring of \coo\ is near the edge of the millimeter disk, typically taken to be 60 AU \citep{Andrews12}. A rapid drop in the surface density of millimeter grains could give rise to higher gas temperatures, which is hinted at in our data, and/or increase the flux of photo-desorbsing UV radiation, leading to an increase of gas phase CO \citep{Cleeves16}. Recently, models including increased desorption of CO have been shown to reproduce an outer ring of DCO$^+$ emission, near the edge of the millimeter dust emission disk \citep{Oberg15}. If chemical processes involving CO can produce rings of emission in DCO$^+$,  it is reasonable to expect CO emission rings as well.

\section{Summary}\label{summary}
We have presented resolved ALMA observations of the \coo\ 3-2, \cooo\ 3-2, \coo\ 6-5, and \cooo\ 6-5 line emission towards the transition disk TW Hya.
Using these observations we construct a radial gas temperature profile, which provides an observational upper limit on the CO freeze-out temperature of $<21$ K.
Using this temperature profile, along with the previous detection of HD 1-0 in this system, we calculate the surface density of the warm gas along with the radial CO abundance relative to \hh. We find that the surface density of the warm gas mass as traced by HD is $\Sigma_{warm\,gas} = 4.7^{+3.0} _{-2.9} \mathrm{\,g\, cm^{-2}}\left(R/10\,\mathrm{AU}\right)^{-1/2}$. The CO abundance is uniformly of order \eten{-6}, failing to return to ISM values in the range $R=10-60$ AU. This, combined with the low abundances of other carbon bearing species in this system, suggests that the majority of the volatile carbon in TW Hya has been removed from the gas.

The ALMA data provide a measurement of the surface CO snowline at $R\sim30$~AU, and allow us to calculate the radius of the midplane snowline. 
Using our CO surface density and temperature profiles to constrain the midplane density of gas phase CO and the CO binding energy respectively, as well as the model midplane gas temperature structure of \citet{Cleeves15}, we expect the midplane CO snowline to occur between 17-23~AU. 
The \coo\ 3-2 and \cooo\ 3-2 emission also show evidence of an outer ring of emission with a minimum at $R\sim36$~AU and a secondary peak at $R\sim52$~AU.
\\

This paper makes use of the following ALMA data: ADS/JAO.ALMA\#2012.1.00422.S. ALMA is a partnership of ESO (representing its member states), NSF (USA) and NINS (Japan), together with NRC (Canada) and NSC and ASIAA (Taiwan) and KASI (Republic of Korea), in cooperation with the Republic of Chile. The Joint ALMA Observatory is operated by ESO, AUI/NRAO and NAOJ.
This work was supported by funding from the National Science Foundation grant AST-1514670 and AST-1344133 (INSPIRE).


\begin{thebibliography}{48}
\expandafter\ifx\csname natexlab\endcsname\relax\def\natexlab#1{#1}\fi

\bibitem[{{Aikawa} \& {Herbst}(1999)}]{Aikawa97}
{Aikawa}, Y., \& {Herbst}, E. 1999, \aap, 351, 233

\bibitem[{{Aikawa} {et~al.}(2002){Aikawa}, {van Zadelhoff}, {van Dishoeck}, \&
  {Herbst}}]{Aikawa02}
{Aikawa}, Y., {van Zadelhoff}, G.~J., {van Dishoeck}, E.~F., \& {Herbst}, E.
  2002, \aap, 386, 622

\bibitem[{{Andrews} {et~al.}(2012){Andrews}, {Wilner}, {Hughes}, {Qi},
  {Rosenfeld}, {{\"O}berg}, {Birnstiel}, {Espaillat}, {Cieza}, {Williams},
  {Lin}, \& {Ho}}]{Andrews12}
{Andrews}, S.~M., {Wilner}, D.~J., {Hughes}, A.~M., {et~al.} 2012, \apj, 744,
  162

\bibitem[{{Barrado Y Navascu{\'e}s}(2006)}]{Barrado06}
{Barrado Y Navascu{\'e}s}, D. 2006, \aap, 459, 511

\bibitem[{{Bergin} {et~al.}(2014){Bergin}, {Cleeves}, {Crockett}, \&
  {Blake}}]{Bergin14}
{Bergin}, E.~A., {Cleeves}, L.~I., {Crockett}, N., \& {Blake}, G.~A. 2014,
  Faraday Discussions, 168, 61

\bibitem[{{Bergin} {et~al.}(2013){Bergin}, {Cleeves}, {Gorti}, {Zhang},
  {Blake}, {Green}, {Andrews}, {Evans}, {Henning}, {{\"O}berg}, {Pontoppidan},
  {Qi}, {Salyk}, \& {van Dishoeck}}]{Bergin13}
{Bergin}, E.~A., {Cleeves}, L.~I., {Gorti}, U., {et~al.} 2013, \nat, 493, 644

\bibitem[{{Bruderer} {et~al.}(2012){Bruderer}, {van Dishoeck}, {Doty}, \&
  {Herczeg}}]{Bruderer12}
{Bruderer}, S., {van Dishoeck}, E.~F., {Doty}, S.~D., \& {Herczeg}, G.~J. 2012,
  \aap, 541, A91

\bibitem[{{Calvet} {et~al.}(2002){Calvet}, {D'Alessio}, {Hartmann}, {Wilner},
  {Walsh}, \& {Sitko}}]{Calvet02}
{Calvet}, N., {D'Alessio}, P., {Hartmann}, L., {et~al.} 2002, \apj, 568, 1008

\bibitem[{{Chapillon} {et~al.}(2008){Chapillon}, {Guilloteau}, {Dutrey}, \&
  {Pi{\'e}tu}}]{Chapillon08}
{Chapillon}, E., {Guilloteau}, S., {Dutrey}, A., \& {Pi{\'e}tu}, V. 2008, \aap,
  488, 565

\bibitem[{{Cleeves}(2016)}]{Cleeves16}
{Cleeves}, L.~I. 2016, \apjl, 816, L21

\bibitem[{{Cleeves} {et~al.}(2014){Cleeves}, {Bergin}, {Alexander}, {Du},
  {Graninger}, {{\"O}berg}, \& {Harries}}]{Cleeves14}
{Cleeves}, L.~I., {Bergin}, E.~A., {Alexander}, C.~M.~O., {et~al.}
  2014, Science, 345, 1590

\bibitem[{{Cleeves} {et~al.}(2015){Cleeves}, {Bergin}, {Qi}, {Adams}, \&
  {{\"O}berg}}]{Cleeves15}
{Cleeves}, L.~I., {Bergin}, E.~A., {Qi}, C., {Adams}, F.~C., \& {{\"O}berg},
  K.~I. 2015, \apj, 799, 204

\bibitem[{{Collings} {et~al.}(2003){Collings}, {Dever}, {Fraser}, {McCoustra},
  \& {Williams}}]{Collings03}
{Collings}, M.~P., {Dever}, J.~W., {Fraser}, H.~J., {McCoustra}, M.~R.~S., \&
  {Williams}, D.~A. 2003, \apj, 583, 1058

\bibitem[{{Dartois} {et~al.}(2003){Dartois}, {Dutrey}, \&
  {Guilloteau}}]{Dartois03}
{Dartois}, E., {Dutrey}, A., \& {Guilloteau}, S. 2003, \aap, 399, 773

\bibitem[{{Dent} {et~al.}(2013){Dent}, {Thi}, {Kamp}, {Williams}, {Menard},
  {Andrews}, {Ardila}, {Aresu}, {Augereau}, {Barrado y Navascues}, {Brittain},
  {Carmona}, {Ciardi}, {Danchi}, {Donaldson}, {Duchene}, {Eiroa}, {Fedele},
  {Grady}, {de Gregorio-Molsalvo}, {Howard}, {Hu{\'e}lamo}, {Krivov},
  {Lebreton}, {Liseau}, {Martin-Zaidi}, {Mathews}, {Meeus},
  {Mendigut{\'{\i}}a}, {Montesinos}, {Morales-Calderon}, {Mora}, {Nomura},
  {Pantin}, {Pascucci}, {Phillips}, {Pinte}, {Podio}, {Ramsay}, {Riaz},
  {Riviere-Marichalar}, {Roberge}, {Sandell}, {Solano}, {Tilling}, {Torrelles},
  {Vandenbusche}, {Vicente}, {White}, \& {Woitke}}]{Dent13}
{Dent}, W.~R.~F., {Thi}, W.~F., {Kamp}, I., {et~al.} 2013, \pasp, 125, 477

\bibitem[{{Du} {et~al.}(2015){Du}, {Bergin}, \& {Hogerheijde}}]{Du15}
{Du}, F., {Bergin}, E.~A., \& {Hogerheijde}, M.~R. 2015, \apjl, 807, L32

\bibitem[{{Dutrey} {et~al.}(2003){Dutrey}, {Guilloteau}, \& {Simon}}]{Dutrey03}
{Dutrey}, A., {Guilloteau}, S., \& {Simon}, M. 2003, \aap, 402, 1003

\bibitem[{{Favre} {et~al.}(2013){Favre}, {Cleeves}, {Bergin}, {Qi}, \&
  {Blake}}]{Favre13}
{Favre}, C., {Cleeves}, L.~I., {Bergin}, E.~A., {Qi}, C., \& {Blake}, G.~A.
  2013, \apjl, 776, L38

\bibitem[{{Fedele} {et~al.}(2013){Fedele}, {Bruderer}, {van Dishoeck},
  {Hogerheijde}, {Panic}, {Brown}, \& {Henning}}]{Fedele13}
{Fedele}, D., {Bruderer}, S., {van Dishoeck}, E.~F., {et~al.} 2013, \apjl, 776,
  L3

\bibitem[{{Gorti} {et~al.}(2011){Gorti}, {Hollenbach}, {Najita}, \&
  {Pascucci}}]{Gorti11}
{Gorti}, U., {Hollenbach}, D., {Najita}, J., \& {Pascucci}, I. 2011, \apj, 735,
  90

\bibitem[{{Guilloteau} {et~al.}(2011){Guilloteau}, {Dutrey}, {Pi{\'e}tu}, \&
  {Boehler}}]{Guilloteau11}
{Guilloteau}, S., {Dutrey}, A., {Pi{\'e}tu}, V., \& {Boehler}, Y. 2011, \aap,
  529, A105

\bibitem[{{Hollenbach} {et~al.}(2009){Hollenbach}, {Kaufman}, {Bergin}, \&
  {Melnick}}]{Hollenbach09}
{Hollenbach}, D., {Kaufman}, M.~J., {Bergin}, E.~A., \& {Melnick}, G.~J. 2009,
  \apj, 690, 1497

\bibitem[{{Kama} {et~al.}(2016){Kama}, {Bruderer}, {Carney}, {Hogerheijde},
  {van Dishoeck}, {Fedele}, {Baryshev}, {Boland}, {G{\"u}sten}, {Aikutalp},
  {Choi}, {Endo}, {Frieswijk}, {Karska}, {Klaassen}, {Koumpia}, {Kristensen},
  {Leurini}, {Nagy}, {Perez Beaupuits}, {Risacher}, {van der Marel}, {van
  Kempen}, {van Weeren}, {Wyrowski}, \& {Y{\i}ld{\i}z}}]{Kama16}
{Kama}, M., {Bruderer}, S., {Carney}, M., {et~al.} 2016, ArXiv e-prints,
  arXiv:1601.01449

\bibitem[{{Kastner} {et~al.}(2015){Kastner}, {Qi}, {Gorti}, {Hily-Blant},
  {Oberg}, {Forveille}, {Andrews}, \& {Wilner}}]{Kastner15}
{Kastner}, J.~H., {Qi}, C., {Gorti}, U., {et~al.} 2015, \apj, 806, 75

\bibitem[{{Lacy} {et~al.}(1994){Lacy}, {Knacke}, {Geballe}, \&
  {Tokunaga}}]{Lacy94}
{Lacy}, J.~H., {Knacke}, R., {Geballe}, T.~R., \& {Tokunaga}, A.~T. 1994,
  \apjl, 428, L69

\bibitem[{{Langer} {et~al.}(2014){Langer}, {Velusamy}, {Pineda}, {Willacy}, \&
  {Goldsmith}}]{Langer14}
{Langer}, W.~D., {Velusamy}, T., {Pineda}, J.~L., {Willacy}, K., \&
  {Goldsmith}, P.~F. 2014, \aap, 561, A122

\bibitem[{{Linsky}(1998)}]{Linsky98}
{Linsky}, J.~L. 1998, \ssr, 84, 285

\bibitem[{{Menu} {et~al.}(2014){Menu}, {van Boekel}, {Henning}, {Chandler},
  {Linz}, {Benisty}, {Lacour}, {Min}, {Waelkens}, {Andrews}, {Calvet},
  {Carpenter}, {Corder}, {Deller}, {Greaves}, {Harris}, {Isella}, {Kwon},
  {Lazio}, {Le Bouquin}, {M{\'e}nard}, {Mundy}, {P{\'e}rez}, {Ricci},
  {Sargent}, {Storm}, {Testi}, \& {Wilner}}]{Menu14}
{Menu}, J., {van Boekel}, R., {Henning}, T., {et~al.} 2014, \aap, 564, A93

\bibitem[{{Miotello} {et~al.}(2014){Miotello}, {Bruderer}, \& {van
  Dishoeck}}]{Miotello14}
{Miotello}, A., {Bruderer}, S., \& {van Dishoeck}, E.~F. 2014, \aap, 572, A96

\bibitem[{{Nomura} {et~al.}(2015){Nomura}, {Tsukagoshi}, {Kawabe}, {Ishimoto},
  {Okuzumi}, {Muto}, {Kanagawa}, {Ida}, {Walsh}, {Millar}, \& {Bai}}]{Nomura15}
{Nomura}, H., {Tsukagoshi}, T., {Kawabe}, R., {et~al.} 2015, ArXiv e-prints,
  arXiv:1512.05440

\bibitem[{{{\"O}berg} {et~al.}(2015){{\"O}berg}, {Furuya}, {Loomis}, {Aikawa},
  {Andrews}, {Qi}, {van Dishoeck}, \& {Wilner}}]{Oberg15}
{{\"O}berg}, K.~I., {Furuya}, K., {Loomis}, R., {et~al.} 2015, \apj, 810, 112

\bibitem[{{{\"O}berg} {et~al.}(2005){{\"O}berg}, {van Broekhuizen}, {Fraser},
  {Bisschop}, {van Dishoeck}, \& {Schlemmer}}]{Oberg05}
{{\"O}berg}, K.~I., {van Broekhuizen}, F., {Fraser}, H.~J., {et~al.} 2005,
  \apjl, 621, L33

\bibitem[{{Penzias} {et~al.}(1972){Penzias}, {Solomon}, {Jefferts}, \&
  {Wilson}}]{Penzias72}
{Penzias}, A.~A., {Solomon}, P.~M., {Jefferts}, K.~B., \& {Wilson}, R.~W. 1972,
  \apjl, 174, L43

\bibitem[{{Qi} {et~al.}(2015){Qi}, {{\"O}berg}, {Andrews}, {Wilner}, {Bergin},
  {Hughes}, {Hogherheijde}, \& {D'Alessio}}]{Qi15}
{Qi}, C., {{\"O}berg}, K.~I., {Andrews}, S.~M., {et~al.} 2015, \apj, 813, 128

\bibitem[{{Qi} {et~al.}(2004){Qi}, {Ho}, {Wilner}, {Takakuwa}, {Hirano},
  {Ohashi}, {Bourke}, {Zhang}, {Blake}, {Hogerheijde}, {Saito}, {Choi}, \&
  {Yang}}]{Qi04}
{Qi}, C., {Ho}, P.~T.~P., {Wilner}, D.~J., {et~al.} 2004, \apjl, 616, L11

\bibitem[{{Qi} {et~al.}(2013){Qi}, {{\"O}berg}, {Wilner}, {D'Alessio},
  {Bergin}, {Andrews}, {Blake}, {Hogerheijde}, \& {van Dishoeck}}]{Qi13b}
{Qi}, C., {{\"O}berg}, K.~I., {Wilner}, D.~J., {et~al.} 2013, Science, 341, 630

\bibitem[{{Reboussin} {et~al.}(2015){Reboussin}, {Wakelam}, {Guilloteau},
  {Hersant}, \& {Dutrey}}]{Reboussin15}
{Reboussin}, L., {Wakelam}, V., {Guilloteau}, S., {Hersant}, F., \& {Dutrey},
  A. 2015, \aap, 579, A82

\bibitem[{{Ros} \& {Johansen}(2013)}]{Ros13}
{Ros}, K., \& {Johansen}, A. 2013, \aap, 552, A137

\bibitem[{{Rosenfeld} {et~al.}(2012){Rosenfeld}, {Qi}, {Andrews}, {Wilner},
  {Corder}, {Dullemond}, {Lin}, {Hughes}, {D'Alessio}, \& {Ho}}]{Rosenfeld12}
{Rosenfeld}, K.~A., {Qi}, C., {Andrews}, S.~M., {et~al.} 2012, \apj, 757, 129

\bibitem[{{Thi} {et~al.}(2010){Thi}, {Mathews}, {M{\'e}nard}, {Woitke},
  {Meeus}, {Riviere-Marichalar}, {Pinte}, {Howard}, {Roberge}, {Sandell},
  {Pascucci}, {Riaz}, {Grady}, {Dent}, {Kamp}, {Duch{\^e}ne}, {Augereau},
  {Pantin}, {Vandenbussche}, {Tilling}, {Williams}, {Eiroa}, {Barrado},
  {Alacid}, {Andrews}, {Ardila}, {Aresu}, {Brittain}, {Ciardi}, {Danchi},
  {Fedele}, {de Gregorio-Monsalvo}, {Heras}, {Huelamo}, {Krivov}, {Lebreton},
  {Liseau}, {Martin-Zaidi}, {Mendigut{\'{\i}}a}, {Montesinos}, {Mora},
  {Morales-Calderon}, {Nomura}, {Phillips}, {Podio}, {Poelman}, {Ramsay},
  {Rice}, {Solano}, {Walker}, {White}, \& {Wright}}]{Thi10}
{Thi}, W.-F., {Mathews}, G., {M{\'e}nard}, F., {et~al.} 2010, \aap, 518, L125

\bibitem[{{Vacca} \& {Sandell}(2011)}]{Vacca11}
{Vacca}, W.~D., \& {Sandell}, G. 2011, \apj, 732, 8

\bibitem[{{van Dishoeck} \& {Black}(1988)}]{vanDishoeck88}
{van Dishoeck}, E.~F., \& {Black}, J.~H. 1988, \apj, 334, 771

\bibitem[{{van Zadelhoff} {et~al.}(2001){van Zadelhoff}, {van Dishoeck}, {Thi},
  \& {Blake}}]{vanZadelhoff01}
{van Zadelhoff}, G.-J., {van Dishoeck}, E.~F., {Thi}, W.-F., \& {Blake}, G.~A.
  2001, \aap, 377, 566

\bibitem[{{Visser} {et~al.}(2009){Visser}, {van Dishoeck}, \&
  {Black}}]{Visser09}
{Visser}, R., {van Dishoeck}, E.~F., \& {Black}, J.~H. 2009, \aap, 503, 323

\bibitem[{{Wichmann} {et~al.}(1998){Wichmann}, {Bastian}, {Krautter},
  {Jankovics}, \& {Rucinski}}]{Wichmann98}
{Wichmann}, R., {Bastian}, U., {Krautter}, J., {Jankovics}, I., \& {Rucinski},
  S.~M. 1998, \mnras, 301, 39L

\bibitem[{{Williams} \& {Cieza}(2011)}]{Williams11}
{Williams}, J.~P., \& {Cieza}, L.~A. 2011, \araa, 49, 67

\bibitem[{{Wilson}(1999)}]{Wilson99}
{Wilson}, T.~L. 1999, Reports on Progress in Physics, 62, 143

\bibitem[{{Zhang} {et~al.}(2016){Zhang}, {Bergin}, {Blake}, {Cleeves},
  {Hogerheijde}, {Salinas}, \& {Schwarz}}]{Zhang16}
{Zhang}, K., {Bergin}, E.~A., {Blake}, G.~A., {et~al.} 2016, \apjl, 818, L16

\end{thebibliography}
\end{document}